\documentstyle[11pt,aaspp4]{article}

\def\etal{et\ al.}
\def\CIV{C~{\sc iv}}
\def\MgII{Mg~{\sc ii}}
\def\HI{H~{\sc i}}
\def\MgI{Mg~{\sc i}}

\def\hMpc{$h^{-1}$~Mpc}
\def\kms{km~s$^{-1}$}

\lefthead{QUASHNOCK \& VANDEN BERK}
\righthead{FORM AND EVOLUTION OF ABSORBER CLUSTERING}

\begin{document}

\title{The Form and Evolution of the Clustering of QSO 
		Heavy--Element Absorption--Line Systems}

\author{Jean M. Quashnock} 
\affil{University of Chicago, Department of Astronomy and Astrophysics,
 5640 South Ellis, Chicago, IL 60637}
\and
\author{Daniel E. Vanden Berk\altaffilmark{1}}
\affil{University of Texas, McDonald Observatory, 
 RLM 15.308, Austin, TX 78712-1083;\\
jmq@oddjob.uchicago.edu, danvb@astro.as.utexas.edu}
\altaffiltext{1}{Harlan J. Smith Postdoctoral Fellow}

\received{23 June 1997}
\revised{11 December 1997}
\accepted{22 January 1998}

\slugcomment{To appear in {\em The Astrophysical Journal}, vol. 500}

\begin{abstract}
We have analyzed the clustering of \CIV\ and \MgII\
absorption--line systems on comoving scales $r$ from 1 to 16 \hMpc ,
using an extensive catalog of heavy--element QSO absorbers
with mean redshift $\langle z\rangle_{\rm \CIV} = 2.2$ 
and $\langle z\rangle_{\rm \MgII} = 0.9$.
We find that, for the \CIV\ sample as a whole,
the absorber line--of--sight correlation function 
is well--fit by a power law of the form
$\xi_{\rm aa}(r)={\left(r_0/r\right)}^\gamma$,
with maximum--likelihood values of $\gamma = 1.75\,^{+0.50}_{-0.70}$
and comoving $r_0 = 3.4\,^{+0.7}_{-1.0}$ \hMpc\ ($q_0=0.5$).
The clustering of absorbers at high redshift is thus 
of a {\em form} that is consistent with that found for galaxies and clusters
at low redshift, and of amplitude such that absorbers are
correlated on scales of galaxy clusters.
We also trace the {\em evolution} of the mean amplitude 
$\xi_0(z)$ of the correlation function,
as a function of redshift, from $z=3$ to $z=0.9$.
We find that, when parametrized in the conventional manner as 
$\xi_0(z)\propto (1+z)^{-(3+\epsilon)+\gamma}$,
the amplitude grows 
with decreasing redshift, with maximum--likelihood value for the
evolutionary parameter of $\epsilon = 2.05 \pm 1.0 $ ($q_0=0.5$).
When extrapolated to zero redshift, the amplitude of the
correlation function implies that the
correlation length $r_0 = 30\,^{+22}_{-13}$ \hMpc\ ($q_0=0.5$) .
This suggests that strong \CIV\ and \MgII\ absorbers, on megaparsec scales,
are biased tracers of the higher--density regions of space,
and that agglomerations of strong absorbers
along a line of sight are indicators of clusters and superclusters.
This is supported by recent observations of ``Lyman break'' galaxies.
The growth seen in the clustering of absorbers
is consistent with gravitationally induced growth of perturbations.
\end{abstract}


\section{Introduction}

The exact nature of QSO heavy--element absorption--line systems
is not yet well understood. Nevertheless, there is growing
evidence suggesting that these absorbers are
associated with galaxy halos and disks.
Deep images of fields around QSOs 
with absorbers in their spectra have been made,
and galaxies have been found at the redshifts 
and in the vicinities of absorbers (see, e.g., \cite{Steidel94};
\cite{Steidel97}).
The distances between galaxies and the associated absorbers 
are inferred to be of order 100 kpc,
or the size of an extended galaxy halo (\cite{Churchill96}).
This association has also been inferred for a substantial fraction
of the Ly$\alpha$ absorbers (\cite{Lanz95}; \cite{LeBrun96}),
but they are not the subject of this work.

One approach in relating absorbers to galaxies is to 
compare their respective clustering properties.
In a previous paper, Quashnock, Vanden Berk, \& York 
(1996, hereafter QVY)
analyzed line--of--sight correlations of 
\CIV\ and \MgII\ absorption--line systems on large scales,
using an extensive catalog of 2200 
heavy--element absorption--line systems in over 500 QSO spectra (\cite{Y98}).
(More details about the catalog 
can be found in an earlier version of the catalog [\cite{Y91}],
as well as in QVY.)
These authors found that the absorbers are clustered, like galaxies,
on comoving scales of 50 to 100 \hMpc\ ($q_0=0.5$).

This high--redshift ($z$ from 1.5 to 3.5) superclustering
of the absorbers is on the same comoving scale as that
traced by the voids and walls of galaxy redshift surveys of 
the local universe (see, e.g., \cite{Kir81}; \cite{GH89}; \cite{Landy96}).
It thus appears that the absorbers are tracing the same 
structure as that traced by galaxies on very large scales,
and that they are effective probes of very large scale structure
in the universe (see, e.g., \cite{Crotts85}; \cite{Tytler93}).
The authors of QVY argued that this superclustering is generic
(they found potential superclusters along 7 different lines of sight);
indeed, such superclustering has been detected by many studies
(see, e.g., \cite{HHW89}; \cite{Dinshaw96}; \cite{Williger96};
and refs. in QVY).

In this paper, we extend the QVY analysis of
line--of--sight correlations of \CIV\ and \MgII\ absorption--line systems
to smaller comoving scales --- from 1 to 16 \hMpc , i.e., 
galaxy cluster scales,
corresponding to line--of --sight velocity differences
$\Delta v \sim $ 200 -- 3000 \kms\ 
at $\langle z\rangle_{\rm \CIV} = 2.2$ ---
and relate the small--scale clustering of absorbers 
to galaxy clustering in general.

Early studies (\cite{Young82}; \cite{SBS88}),
with approximately 100 \kms\ spectral resolution,
found that \CIV\ absorbers cluster significantly on these scales,
with a correlation function $\xi \sim $ 5 to 10
for line--of--sight velocity differences $\Delta v =$ 200 -- 600 \kms .
On smaller scales ($\Delta v < 200$ \kms),
internal motions of the gas comprising the absorption systems 
are expected to dominate (Heisler \etal\ 1989),
if the absorbing gas is moving in gravitational orbits inside galactic halos
with typical cloud velocity dispersions $\sim 200$ \kms (\cite{DP83}).
Indeed, Petitjean \& Bergeron (1994), using higher spectral resolution data,
found that the \CIV\ line--of--sight correlation function
can be described by two Gaussian components, with velocity
widths $\sigma_v =$ 109 and 525 \kms,
with the latter component accounting for almost all (93\%) of the signal
in $\xi$ for $\Delta v =$ 200 -- 600 \kms .
A similar result, with even higher resolution Keck High Resolution
Echelle Spectrometer (HIRES) spectra,
has been found by Rauch \etal\ (1996).

A natural interpretation (see, e.g., Sargent \etal\ 1988) of these results 
is that the low--velocity component ($\Delta v <$ 200 \kms)
is due to gas motion inside the absorbers,
with the higher--velocity component ($\Delta v >$ 200 \kms)
being due to actual spatial clustering
of the absorbers on galaxy cluster scales (several Mpc).
However, because the line--of--sight velocity dispersion 
of typical galaxy clusters ranges from 300 -- 1400 \kms (\cite{Zab93};
\cite{Dinshaw96}), the higher--velocity component may represent,
at least in part, motion of absorbers inside a galaxy cluster,
rather than true spatial clustering (\cite{Shi}). This has been argued by
Crotts, Burles, \& Tytler (1997), who explore the spatial clustering
of \CIV\ systems along adjacent lines of sight, and claim that it is
significantly weaker than clustering along a line of sight.

We return to the question of interpretation of the clustering
of heavy--element absorption lines in \S\ 6. 
Our interest here is in the {\em form} and {\em evolution}
of that clustering on comoving scales of 1 \hMpc\ and greater
($\Delta v > $ 180 \kms), corresponding to the higher--velocity
component. We have taken steps (\S\ 3) to insure that there is no aliasing of
power from the low--velocity component.

The outline of the paper is as follows.
We present our likelihood method of estimating
the correlation function from the data in \S\ 2.
We describe our data selection criteria in \S\ 3.
We then discuss the {\em form} of the correlation function in \S\ 4.
We also present results on the {\em evolution} of the clustering,
by calculating its redshift dependence in \S\ 5.
We discuss some of the implications of these results in \S\ 6, 
and summarize them in \S\ 7.
In the Appendix, we explicitly calculate the likelihood as a function of data
and model parameters.

\section{Likelihood Method of Estimating the Absorber Correlation Function}

Here we outline our likelihood method 
of estimating the line--of--sight correlation function, $\xi_{\rm aa}(r,z)$.
(We defer a detailed calculation of the likelihood function to the Appendix.)
Unless otherwise noted, we take $q_0=0.5$ and $\Lambda=0$.
We follow the usual convention and take the Hubble constant
to be 100\,$h$\ km~s$^{-1}$~Mpc$^{-1}$.

The correlation function $\xi_{\rm aa}(r,z)$ relates the {\em observed} 
number of absorber pairs,
separated by comoving distance $r$ at redshift $z$, 
with the {\em expected} number of pairs, of the same separation and redshift,
from an unclustered absorber population.
The data $D$ consist of the observed set of absorber pairs $i$,
each with comoving separation $r_i$ at redshift $z_i$.
Normally, in calculating the correlation function, these data
are binned, and $\xi_{\rm aa}(r,z)$ is obtained by dividing the
observed number of pairs in a bin by the number expected in the bin.
In our likelihood approach, no such binning is necessary.

Consider a given model for $\xi_{\rm aa}(r,z)$. 
Each such model is described by a set of parameters $M$,
and we wish to find the parameter values
which best describe the observed data $D$.
To do so, we must maximize the {\em likelihood} $\cal L$, 
(calculated in the Appendix, eq.~[A4]),
which is the probability of the data $D$,
given the model parameters $M$ for
the correlation function $\xi_{\rm aa}(r,z)$: 
${\cal L} \equiv P(D|M)$.
Given a prior distribution, $P(M)$, for the model parameters
(usually obtained from physical or symmetry arguments),
we can use Bayes' Theorem to then calculate the posterior probability
distribution, $P(M|D)$, for the model parameters $M$, given the data $D$
(see \cite{Lor92} for a compendium of astrophysical applications of
Bayesian inference):
\begin{equation}
P(M|D) = P(D|M)\times P(M)/ \int P(D|M) P(M) dM \; .
\end{equation}

The parameter values which maximize the likelihood are
those we take as the ``best--fit'' values,
and credible regions for the parameters around the maximum--likelihood values
are obtained from equation~(1), by requiring that a fiducial amount
of posterior probability (such as 68.3\% or 95.5\%, 
nominal 1 and 2 $\sigma$ amounts) be contained inside the credible region.

In this paper, we consider three models for the correlation function.
The first is a simple power law, with comoving clustering length $r_0$
and power--law index $\gamma$:
\begin{equation}
\xi_{\rm aa}(r,z) = {\left(r_0/r\right)}^\gamma \; .
\end{equation}
The second is also a power law, but with an amplitude 
that evolves with decreasing redshift as a power of the expansion factor:
\begin{equation}
\xi_{\rm aa}(r,z) = {\left(r_0/r\right)}^\gamma 
{\left({1+z\over 1+z_0}\right)}^{-(3+\epsilon)+\gamma}\; .
\end{equation}
Again $r_0$ is the comoving clustering length (at a fixed redshift $z_0$),
$\epsilon$ is the conventional evolutionary parameter 
(\cite{GP77}; \cite{Efstat91}),
and $\gamma$ is a power--law index that does not evolve with redshift.
The third model is that of a power--law correlation function
evolving according to linear theory
of gravitational instability (\cite{Peeb80}, 1993):
\begin{equation}
\xi_{\rm aa}(r,z) = {\left(r_0/r\right)}^\gamma g^2(z,\Omega_0,\Lambda)\; ,
\end{equation}
where $g(z,\Omega_0,\Lambda)$ is the growth factor.
When $\Omega_0=1$ and $\Lambda=0$, $g\propto (1+z)^{-1}$.

\section{Selection of the Data from the Absorber Catalog}

We have drawn our \CIV\ and \MgII\ data sample from the heterogeneous
catalog of Vanden Berk \etal\ (1998) ---
with the aim of constructing as homogeneous a dataset as possible ---
by accounting for the spectral wavelength coverage, spectral
sensitivity, and spectral resolution of each line of sight.  
The selection criteria we used to produce the final set of absorption systems 
are similar to those in QVY, and we outine them here.

First, the Ly$\alpha$ forest region, where heavy--element line
identifications are often unreliable, was excluded in each spectrum.
The so--called associated region, within 5000 \kms\ of each QSO, was
also excluded, because the origin of the systems in that region may be different
than that of systems farther removed from the QSO (\cite{Foltz}).  
Only systems with at least a \CIV\ or \MgII\ doublet were selected; 
in that case, the intrinsically stronger doublet component was required
to have been detected at the $5\sigma$ or greater significance level.
In addition, a system had to have at least one more identified 
heavy--element line, or have a doublet equivalent width ratio 
that is consistent with atomic physics (within the measurement errors). 

Since we are measuring correlations of absorber pairs with comoving separations
of 1 \hMpc\ or greater, we selected only those
spectra or spectral regions in which the spatial resolution was
better than this limit.  All systems lying within 1 \hMpc\ of
each other, in the same line of sight, were combined into a single system
by averaging the redshifts and taking the equivalent width of the strongest
system. This has the effect of minimizing the aliasing of power from
scales smaller than 1 \hMpc\ ---
corresponding to $\Delta v = 180$ \kms\ in the rest frame 
at $\langle z\rangle_{\rm \CIV} = 2.2$ ---
where internal motions of the gas comprising the absorption systems 
are thought to dominate (Heisler \etal\ 1989; and \S\ 1 above).
Since 180 \kms\ is 1.6 times as large as the low--velocity
width (\cite{PB94}, Fig. 4a),
most of the clustering from the low--velocity Gaussian component
has been removed by our selection procedure.

Applying all the selection criteria above to the 
Vanden Berk \etal\ (1998) catalog leaves a data sample 
consisting of 260 \CIV\ absorbers, drawn from 202 lines of sight,
with redshifts ranging from $1.2 < z < 3.6$ and mean redshift 
$\langle z\rangle_{\rm \CIV} = 2.2$, and 64 \MgII\ absorbers,
drawn from 278 lines of sight, with redshifts ranging 
from $0.3 < z < 1.6$ and mean redshift $\langle z\rangle_{\rm \MgII} = 0.9$.

\section{Form of the Correlation Function}

Figure~1 shows the 
line--of--sight correlation function $\xi_{\rm aa}(r)$,
for the entire sample of \CIV\ absorbers
(with mean redshift $\langle z\rangle_{\rm \CIV} = 2.2$),
as a function of absorber comoving separation $r$ from 1 to 16 \hMpc ,
in 4 octaves.
We have restricted our analysis to these scales, because
data on scales larger than 16 \hMpc\ is too scarce 
to be used unless it is put in very large bins, 
and  because we did not want to examine scales
smaller than 1 \hMpc .

The correlation function is computed by comparing
the distribution of real absorber pair separations
with the expected distribution for an unclustered sample.
We calculated the expected distribution using Monte Carlo simulations,
creating a large number of fake unclustered catalogs,
using the following ``bootstrap'' method (see QVY).  
For each Monte Carlo simulation, 
the same number of absorption systems as in the real sample
catalog was chosen for random distribution.  
The redshifts and equivalent
widths of the random systems were kept the same as the real systems, as
were the spectral properties of the observed QSO lines of sight.
The systems were then randomly placed along the lines of sight, 
subject to the condition that a system could have been detected 
in the given line of sight.  This method randomizes the absorber placements, 
while taking into account the redshift and equivalent width number density, 
which remain exactly the same as in the real sample. 
The expected distribution of pair separations 
was then calculated by averaging the pair distribution over
1000 Monte Carlo simulations.
The vertical error bars through the data points 
are 1 $\sigma$ errors in the  estimator for $\xi_{\rm aa}$ (\cite{Peeb80}),
and were computed in the same fashion as in QVY.

The first point in Figure~1 shows $\xi_{\rm aa}(r)$ for $r$ between
1 and 2 \hMpc, corresponding to 180 -- 360 \kms\ at
$\langle z\rangle_{\rm \CIV} = 2.2$.
We find that $\xi_{\rm aa} = 5.4 \pm 1.7$,
a value consistent with several previous determinations:
\noindent
(1) $\xi_{\rm aa} = 5.7 \pm 0.6$
for $\Delta v =$ 200 -- 600 \kms
(sample A2 of Sargent et al. 1988);
\noindent
(2) $\xi_{\rm aa} \sim 6$ at 360 \kms
(Fig. 3 of Petitjean \& Bergeron 1994); and
\noindent
(3) $\xi_{\rm aa} \sim 4$ at 300 \kms  
(Fig. 4 of Rauch \etal\ 1996).
However, our dataset is large enough to measure $\xi_{\rm aa}$ on
bigger scales and to determine the form of the correlation function
more accurately.

From Figure~1, the absorber line--of--sight correlation function
on smaller scales can be described by a power law.
However, some information has been lost because of the
{\em binning} of data required to make the errors in Figure~1
reasonably small. Accordingly, a standard $\chi^2$-fit to the data
is not the most powerful method available.

Instead, we have used the maximum--likelihood method of \S\ 2 
(eq. [A4]) to fit a power law
of the form given in equation (2) to the {\em unbinned} \CIV\ data.
We find maximum--likelihood values for the parameters,
obtaining $r_0=3.4$ \hMpc\ and $\gamma=1.75$.
This fit is shown in Figure~1, and describes the data quite well.

Figure~2 shows the credible region for these parameters.
The cross marks the maximum--likelihood location, and the
heavy and light contours mark the 1 $\sigma$ and 2 $\sigma$
credible regions, respectively.
In calculating the credible regions, 
we have used Bayes' Theorem (eq. [1]) 
and taken uniform priors $P(M)$ in the power law intercepts to Figure 1, 
corresponding
\footnote{Since we have no
{\em a priori} information on the scale of clustering, $r_0$,
nor on its overall amplitude, $\xi_{\rm aa}$(1 \hMpc) = $r_0^\gamma$,
we take uniform priors in $\log r_0$ and
$\log \xi_{\rm aa}$(1 \hMpc) = $\gamma \log r_0$ (the intercepts in Fig.~1).  
Thus the prior probability density $dP \propto 
d(\log r_0) \times d(\gamma \log r_0)$, or, using a Jacobian to change
to our two independent variables of interest, $\gamma$ and $r_0$,
$dP \propto d\gamma \times d(\log^2 r_0)$.}
to uniform priors $P(M)$ in the parameters 
$\log^2 r_0$ and $\gamma$.

We condense the information in Figure~2 
by showing the marginalized posterior distributions
for the parameters $\gamma$ and $r_0$ in Figures 3 and 4, respectively.
The vertical lines mark the 1 $\sigma$ credible regions,
which we find to be $1.05 < \gamma < 2.25$, 
and 2.4 \hMpc\ $ < r_0  < $ 4.1 \hMpc\ ($q_0=0.5$).
Thus, for the \CIV\ sample as a whole,
the absorber correlation function 
is well--fit by a power law of the form
$\xi_{\rm aa}(r)={\left(r_0/r\right)}^\gamma$,
with maximum--likelihood values of $\gamma = 1.75\,^{+0.50}_{-0.70}$
and comoving correlation length 
$r_0 = 3.4\,^{+0.7}_{-1.0}$ \hMpc\ ($q_0=0.5$).

Unfortunately, the \MgII\ sample, with only 64 absorbers
(with mean redshift $\langle z\rangle_{\rm \MgII} = 0.9$)
compared to 260 absorbers in the \CIV\ sample,
is not large enough to adequately model the form of
the correlation function. The data are, at the 95\% confidence level,
consistent with all values of $\gamma > 1.55$. Nevertheless,
the sample is large enough to constrain the comoving correlation
length, if we fix the value of the power--law index at $\gamma=1.75$.
In this case, we find that $r_0 = 8.4 \pm 2.0$ \hMpc .
This larger value --- assuming that \CIV\ clustering
is directly comparable to that of \MgII\ (see \S\ 5 below) --- 
suggests that the amplitude of the correlation function has
grown between redshift $\langle z\rangle_{\rm \CIV} = 2.2$
and redshift $\langle z\rangle_{\rm \MgII} = 0.9$.
In the following section, we find that this is indeed the case,
by investigating the evolution of the correlation function, 
using the {\em entire} sample of \CIV\ and \MgII\ absorbers.

\section{Evolution of the Correlation Function}

We have investigated the evolution of the clustering of absorbers
by dividing the \CIV\ absorber sample into three 
approximately equal redshift sub--samples,
and comparing these to the \MgII\ sample.
Figure~5 shows the mean of the correlation function,
$\xi_0(z)$, averaged over comoving scales 
$r$ from 1 to 16 \hMpc,
for the low ($1.2<z<2.0$), medium ($2.0<z<2.8$), 
and high ($2.8<z<3.6$) redshift \CIV\ sub--samples,
as well as for the \MgII\ sample ($0.3<z<1.6$).
The amplitude of the correlation function is clearly growing
with decreasing redshift.

We have used both the \CIV\ and \MgII\ samples to study the evolution
of heavy--element absorber clustering with redshift,
since these samples cover almost disjoint redshift intervals.
For the following reasons, we believe that it is reasonable to do this.
\noindent
(1) Caulet (1989), in a high--sensitivity survey of \MgII\ and \MgI\
lines found in \CIV\ systems, found that 60\% to 70\% of the \MgII\
population ($1 < z < 2$) have moderate or strong \CIV\ absorption
lines (equivalent width $> 0.3$ \AA , like the bulk of our sample),
the rest having either weak (equivalent width $< 0.3$ \AA) or undetected lines.
Thus, \MgII\ systems can reasonably be expected to trace a similar clustering
distribution as stronger \CIV\ systems, 
and a correlation analysis should yield similar results, 
within the errors of $\xi_0(z)$ in Figure 5.
(In the absorption catalog [\cite{Y98}], 
using loose selection criteria, we find that
about 67\% of the \MgII\ systems have corresponding detected \CIV\
lines when they were accessible.  However, after applying the
stricter selection criteria (\S\ 3) used for the clustering analysis, 
only 7 \MgII/\CIV\  systems remain.  
This is not sufficient to directly compare
the clustering of each species.)
\noindent
(2) The work of Petitjean \& Bergeron (1990, 1994) suggests that the
clustering properties of \CIV\ and \MgII\ may be directly comparable.
\noindent
(3) It seems highly unlikely that the distribution of \CIV\ and \MgII\
systems would be different, in a statistical sense, on scales as large as those
studied here, i.e., 1 \hMpc\ to 16 \hMpc , corresponding to galaxy cluster
and {\em intercluster} scales. That would imply a large--scale segregation
of species.
\noindent
(4) The correlation function measured for \MgII\ systems is consistent
with an extrapolation of the best--fit to the \CIV\ data alone (see below).

We have used the maximum--likelihood formalism of \S\ 2 and the Appendix
to fit the {\em unbinned} \CIV\ and \MgII\ data, 
and describe the evolution of the correlation function,
using the evolving power law, given by equation (3).
We fix $z_0$ at the mean \CIV\ redshift of 2.2,
so that the parameter $r_0$ corresponds to the correlation length
at that redshift, enabling a direct comparison with the results of
\S\ 3 for the \CIV\ sample as a whole (see below).
Since the data for the \MgII\ sample, as well as for the \CIV\ sample
as a whole, are consistent with a constant value of the power--law index
$\gamma$, and since growth of hierarchical clustering
in gravitational instability theory (\cite{Peeb80}, 1993) 
and in numerical simulations
occurs with a nearly constant value of $\gamma$,
we have fixed $\gamma$ at its maximum--likelihood value
of 1.75 in our analysis below, and parametrized the
evolution of the correlation function in terms
of the evolutionary parameter $\epsilon$ and 
the correlation length $r_0$ at redshift $z_0=2.2$.

Figure~6 shows the credible region for these parameters.
The heavy and light contours mark the 1 $\sigma$ and 2 $\sigma$
credible regions, respectively.
The cross marks the maximum--likelihood location, 
which we find to be $\epsilon=2.05$ and $r_0=3.2$ \hMpc .
This fit is the one shown as a solid line in Figure~5.

In order to check our assumption that \MgII\ and \CIV\ clustering
are directly comparable, we also show the maximum--likelihood fit
to the \CIV\ data alone (shown as a dashed line in Fig. 5),
which we find to be $\epsilon=2.3$ and $r_0=3.3$ \hMpc .
The correlation function measured for \MgII\ systems is clearly consistent
with the value extrapolated from \CIV\ at higher redshift.
The error in the estimate of the evolutionary parameter $\epsilon$ is
of course larger if we use the \CIV\ data alone ($\epsilon=2.3\pm 1.5$);
hence, we have included the \MgII\ data in our analysis,
for the reasons given above.

We show the marginalized posterior distributions
for the parameters $r_0$ and $\epsilon$, in Figures 7 and 8, respectively.
The vertical lines mark the 1 $\sigma$ credible regions,
which we find to be 2.6 \hMpc\ $ < r_0 < $ 3.8 \hMpc ,
and $1.05 < \epsilon < 3.05$.
All these results are with $q_0=0.5$.
With $q_0=0.1$, we expect our estimate of $\epsilon$
to decrease from 2.05 to 0.8.

It is interesting to compare the posterior distribution for $r_0$
in Figures 4 and 7. Both have a maximum near the same location
at $r_0 \sim 3.3$ \hMpc , but the distribution in Figure 7
is definitely narrower. This is not surprising, since the evolution
in the amplitude of the correlation function 
(and hence in the correlation length) has smeared the distribution of
$r_0$ in Figure 4. There, $r_0$ represents
the correlation length for the \CIV\ sample as a whole (with mean redshift
$\langle z\rangle_{\rm \CIV} = 2.2$), whereas in Figure 7 we have allowed
for evolution, with $r_0$ representing the correlation length
exactly at $z_0=2.2$. The fact that the distribution in Figure 7
is narrower is more evidence in favor of an 
evolving  correlation function.
With $\epsilon = 2.05 \pm 1.0$, the no--evolution case
($\epsilon = -1.25$) is in fact {\em ruled out} at the 3.3 $\sigma$
(99.95 \%) confidence level.

\section{Discussion}

We have analyzed the clustering of \CIV\ and \MgII\
absorption--line systems on comoving scales $r$ from 1 to 16 \hMpc ,
using a new likelihood method that does {\em not} require the usual binning
used to compute the correlation function.
We find that the {\em form} of the line--of--sight correlation function
is well--described by a power law of the form
$\xi_{\rm aa}(r)={\left(r_0/r\right)}^\gamma$, with maximum--likelihood
values for the power--law index of $\gamma = 1.75\,^{+0.50}_{-0.70}$.
The clustering of absorbers at high redshift is thus 
of a form that is consistent with that found for galaxies and clusters,
at low redshift, on megaparsec scales ($\gamma = 1.77 \pm 0.04$ for
galaxies [\cite{DP83}], 
$\gamma = 2.1 \pm 0.3$ for clusters [\cite{Nichol92}]).
It appears that the absorbers are tracing the large--scale structure
seen in the distribution of galaxies and clusters, and are doing so 
at high redshift. The finding strengthens the case for using absorbers
in probing large--scale structure.

We find that the {\em amplitude} of the correlation function, characterized
by the correlation length $r_0$, is growing with 
decreasing redshift. For the entire \CIV\ absorber sample,
with mean redshift $\langle z\rangle_{\rm \CIV} = 2.2$,
we find that $r_0 = 3.4\,^{+0.7}_{-1.0}$ \hMpc\ ($q_0=0.5$),
whereas for the \MgII\ sample,
with mean redshift $\langle z\rangle_{\rm \MgII} = 0.9$,
we find that $r_0 = 8.4 \pm 2.0$ \hMpc .
The \CIV\ sample is large enough to be divided into several sub--samples;
by using all the data sets, and parametrizing the amplitude
of the correlation function in the usual manner as
$\xi_0(z) \propto (1+z)^{-(3+\epsilon)+\gamma}$,
we find that the growth is reflected in a value for
the evolutionary parameter of $\epsilon = 2.05 \pm 1.0$ ($q_0=0.5$).
 
Evidence for a trend of increasing clustering of Ly$\alpha$
absorbers [with $N({\rm\HI}) > 6.3 \times 10^{13}~{\rm cm}^{-2}$]
with decreasing redshift has been found by Cristiani \etal\ (1997).
Like Crotts (1989), these authors also find a clear trend of increasing
Ly$\alpha$ absorber clustering with increasing column density,
and find that an extrapolation to column densities
typical of heavy--element systems 
[$N({\rm\HI}) >  10^{16}~{\rm cm}^{-2}$]
is consistent with the
clustering observed for \CIV\ absorbers (\cite{PB94}; \cite{SC96}).
Our finding of growth in the clustering
of heavy--element systems with decreasing redshift supports both a
continuity scenario between Ly$\alpha$ and heavy--element systems 
(\cite{Crotts89}; \cite{Tytler95}; \cite{Cowie95}; \cite{Cr97}),
and the common action of gravitational instability.

Fern\'andez-Soto \etal\ (1996) have investigated
the correlation function of  Ly$\alpha$ absorbers
[with $N({\rm\HI}) > 3 \times 10^{14}~{\rm cm}^{-2}$] at high redshift,
by using corresponding \CIV\ systems as tracers.
After comparing the observed correlations at high redshift
to the galaxy correlation length at present (viz.,
$r_0\sim 5.5$ \hMpc), they find that the evolutionary
parameter $\epsilon$ is anywhere between 1.5 and 2.8 (approximate 1 $\sigma$
confidence; see Fig. 4 of Fern\'andez-Soto \etal\ 1996), 
consistent with our finding.
However, their determination, unlike ours, is dependent upon comparing
absorber clustering in the past
with galaxy clustering at present. This is problematic, since
the exact relationship between the stronger Ly$\alpha$ absorbers,
presumed to be halos of protogalaxies,
and the corresponding population of galaxies today is unknown,
and different types of galaxies are known to cluster with differing strengths.
Nevertheless, their determination of Ly$\alpha$ absorber
clustering is broadly consistent with what
is expected for galaxies at high redshift,
and their estimate of the evolutionary parameter $\epsilon$ 
is consistent with ours. 

The growth of the correlation function of heavy--element absorbers, 
as shown in Figure 5 and if extrapolated to zero redshift
using the evolving power law (eq. [3]), 
implies a large value for the present--day value of the correlation length: 
$r_0 = 30\,^{+22}_{-13}$ \hMpc\ ($q_0=0.5$) .
This value is several times larger than the average value
for field galaxies today,
viz., $r_0=5.4 \pm 1.0$ \hMpc\ (\cite{DP83}),
but is comparable to that of clusters,
viz., $r_0=16.4 \pm 4.0$ \hMpc\ (\cite{Nichol92}).
The correlation length of galaxies today is
known to depend on luminosity and type (\cite{LeFevre96}),
ranging up to 7.4 \hMpc\
in the Stromlo-APM redshift survey (\cite{Love95})
and up to 14.5 \hMpc\ in the Second Southern Sky Redshift Survey 
(SSRS2; \cite{Benoist96}),
for the most luminous early--type galaxies.
Comparing all these values of the
correlation length suggests that, on megaparsec scales, 
the strong \CIV\ and \MgII\ absorbers in the Vanden Berk \etal\ (1998)
catalog are tracing the distribution of clusters, 
or of the most luminous super-$L_{*}$ galaxies in those clusters.

Of course, extrapolating our best fit in Figure 5 to zero redshift
is highly uncertain, as reflected in the broad range in our estimate
of $r_0$ today, and is dependent on an assumed power--law form for
the evolution of the correlation function.
This makes a determination of heavy--element absorber clustering at low
redshift all the more important;
however, the sample of low--redshift heavy--element absorption--line systems
(see, e.g., \cite{Bahcall96}; \cite{Vanden97})
is too small at present to be useful in this regard.

It is not yet clear whether the correlation function shown in Figure 1
represents motion of absorbers inside a galaxy cluster 
(\cite{Shi}; \cite{Crotts97}),
rather than true spatial clustering, since along a line of sight we cannot
distinguish peculiar velocities from the Hubble flow. In fact, on the scales
considered here, the same ambiguity would arise with galaxies themselves.
What is clear, however, is that the {\em scale} ($r_0=$ 3.4 \hMpc,
corresponding to $\Delta v = $ 610 \kms\ 
at $\langle z\rangle_{\rm \CIV} = 2.2$), 
the {\em form} ($\xi\propto r^{-1.75}$),
and the {\em amplitude} of the clustering in Figure 1 all are indicative
of an association of strong absorbers with clusters.
Furthermore, this clustering is growing with time (Fig.~5),
and could represent actual growth in spatial clustering 
(see below) or an increase in the velocity of absorber motions inside
clusters due to increasing cluster masses.

If heavy--element absorbers are indicative of galaxy clusters,
how do their number densities compare?
The number of strong heavy--element absorption--line systems
per unit redshift along a line of sight is 
$d{\cal N}/dz \gtrsim 1$ (see, e.g., \cite{VQYY96}). 
Taking $\bar n$ and $\bar \sigma$ to be, respectively,
the characteristic comoving number density and effective cross section
of the absorbers, 
and defining $d{\cal N}/dR$ to be the number of systems per comoving length,
we obtain (\cite{Murdoch}, $q_0=0.5$):
\begin{equation}
\bar n \bar \sigma = {d{\cal N}\over dR} = {d{\cal N}\over dz}\times
{H_0\over c} \left(1+z\right)^{3\over 2} \; .
\end{equation}
Taking for an upper limit of $\bar \sigma$
a cross section characteristic of clusters 
having size an Abell radius
$r_a = 1.5$ \hMpc, viz. $\bar \sigma < \pi r_a^2 = 7 h^{-2}$ Mpc$^2$,
we find that $\bar n > 10^{-4} h^3$ Mpc$^{-3}$ at $z\gtrsim 1$
(\cite{Crotts89}),
which is at least twice the density of richness class $R\geq 0$ clusters
found by Postman \etal\ (1996). This means there must be, {\em on average},
several strong heavy--element absorbers per cluster in space.

Then, if the strong clustering seen in the line--of--sight correlation function 
of these absorbers is indicative of clustering on galaxy cluster scales,
they must be {\em biased} tracers of higher density regions of space.
This suggests that agglomerations of strong absorbers
along a line of sight are indicators of clusters and superclusters (see QVY).
This relationship is supported by the following.
\noindent
(1) Steidel \etal\ (1998) have recently discovered
a large concentration of 15 ``Lyman break'' galaxies, 
at redshift $z = 3.09 \pm 0.02$,  of size 11\arcmin\ by 8\arcmin\ 
on the plane of the sky, corresponding to 
10 \hMpc\ by 7 \hMpc\ (comoving, $\Omega_0 = 1$). 
In studying the spectrum of a backlighting QSO ($z_{\rm em} = 3.356$) 
that was also discovered in the field, they found a \CIV\ doublet, 
as well as a strong Ly$\alpha$ system, at $z=3.094$, 
right at the redshift of the concentration.
(Two other prominent features, albeit of lower significance,
were also seen in the redshift histogram of the Lyman break galaxies,
and both have counterparts in metal line systems.)
A similar coherence between the large--scale distribution of high--redshift
galaxies ($z=2.38$) and that of absorbing gas 
has also been reported by Francis \etal\ (1996).
\noindent
(2) Giavalisco \etal\ (1998) have estimated the amplitude of the correlation
function of the aforementioned Lyman break galaxies at a 
median redshift $z = 3.04$,
and find a correlation length $r_0=2.1\pm 0.7$ \hMpc\ ($q_0=0.5$),
a value consistent with that we have found for the \CIV\
absorbers in Figure 5 ($r_0=2.2$ \hMpc\ at $z=3.04$).

In a detailed three--dimensional numerical investigation of the evolution
(from $z=5$ to $z=2$)
of structure in the distribution of Ly$\alpha$ absorbers,
Zhang \etal\ (1998) have found that the richest absorbers 
[$N({\rm\HI}) > 10^{15}~{\rm cm}^{-2}$],
and, correspondingly, heavy--element systems,
tend to concentrate in nodules
at the intersection of filamentary structures (with comoving separation
of a few megaparsecs at $z=2$) comprising the clustering network.
Those nodules appear to represent the locus of galaxy clusters.
The scale of clustering found in our study of heavy--element
absorption--line systems ($r_0=3.2$ \hMpc\ at $z=2.2$)
appears consistent with this picture,
and suggests that these systems are indeed tracing the richer
agglomerations of the clustering network. 
The previous finding by QVY of superclustering of
\CIV\ absorbers on 100 \hMpc\ scales suggests the existence
of clustering in the nodules as well.

The strong clustering that we find 
in the heavy--element absorption--line systems is
also not surprising, given that most of the sample
consists of the strongest systems with relatively large
equivalent widths (order 0.4 \AA\ and greater),
and the recent claims (\cite{Cr97}; \cite{Dod98}) of a strong dependence
of clustering strength on the column density of the systems.
The median \CIV\ $\lambda 1548$ rest equivalent width in our sample is
$\langle W\rangle = 0.4$ {\AA}, and there are few systems with $W<0.1$ {\AA}.
We have divided the sample in half by equivalent width, and computed
the mean correlation function $\xi_0$ averaged for comoving scales 
between 1 \hMpc\ and 16 \hMpc.  We find that
the weaker systems appear to be less clustered than the stronger systems:
$\xi_0(W<0.4\, {\rm\AA}) = 0.87\pm 0.45$,  
$\xi_0(W>0.4\, {\rm\AA}) = 1.73\pm 0.60$; 
however, the sample size is not sufficient to
significantly test the dependence of clustering on equivalent width.

The growth in the correlation function
and the value of the evolutionary parameter 
($\epsilon = 2.05 \pm 1.0$) that we find,
are consistent with gravitationally induced growth of perturbations.
For $q_0=0.5$, from linear theory of gravitational instability
(\cite{Peeb80}, 1993) $\xi \propto {\left(1+z\right)}^{-2}$,
i.e. $\epsilon = 0.75$ if $\gamma=1.75$.
Also, from numerical simulations (\cite{Melott92};
\cite{Carlberg97}; \cite{Colin97}), $\epsilon = 1.0 \pm 0.1$.

It is possible (eq. [4]) to examine
the evolution of the correlation function,
using the growth factor $g(z,\Omega_0,\Lambda)$ from linear theory
of gravitational instability (\cite{Peeb80}, 1993);
however, because of the uncertainties in our determination of the
evolutionary parameter $\epsilon$ and our lack of
understanding of the exact nature of the heavy--element absorbers,
it is not possible to derive values for the cosmological
parameters $\Omega_0$ and $\Lambda$ from this work.
A fruitful approach in the future may be more
detailed numerical simulations of the heavy--element absorber network,
like those of Zhang \etal\ (1998) for the Ly$\alpha$ forest.
In addition, results from the upcoming
Sloan Digital Sky Survey will add greatly to our
understanding of the absorbers.

\section{Summary}

We summarize our results:

1.  We have analyzed the clustering of \CIV\ and \MgII\
absorption--line systems on comoving scales $r$ from 1 to 16 \hMpc ,
using an extensive catalog of heavy--element QSO absorbers
with mean redshift $\langle z\rangle_{\rm \CIV} = 2.2$ 
and $\langle z\rangle_{\rm \MgII} = 0.9$.
We find that, for the \CIV\ sample as a whole,
the absorber line--of--sight correlation function 
is well--fit by a power law of the form
$\xi_{\rm aa}(r)={\left(r_0/r\right)}^\gamma$,
with maximum--likelihood values of $\gamma = 1.75\,^{+0.50}_{-0.70}$
and comoving $r_0 = 3.4\,^{+0.7}_{-1.0}$ \hMpc\ ($q_0=0.5$).

2.  The clustering of absorbers at high redshift is thus 
of a {\em form} that is consistent with that found for galaxies and clusters
at low redshift, and of amplitude such that absorbers are
correlated on scales of clusters of galaxies.
It appears that the absorbers are tracing the large--scale structure
seen in the distribution of galaxies and clusters, and are doing so 
at high redshift. The finding strengthens the case for using absorbers
in probing large--scale structure.

3.  We also trace the {\em evolution} of the mean amplitude 
$\xi_0(z)$ of the correlation function,
as a function of redshift, from $z=3$ to $z=0.9$.
We find that, when parametrized in the conventional manner as 
$\xi_0(z)\propto (1+z)^{-(3+\epsilon)+\gamma}$,
the amplitude grows
with decreasing redshift, with maximum--likelihood value for the
evolutionary parameter of $\epsilon = 2.05 \pm 1.0$ ($q_0=0.5$).

4.  When extrapolated to zero redshift, the amplitude of the
correlation function implies that the
correlation length $r_0 = 30\,^{+22}_{-13}$ \hMpc\ ($q_0=0.5$),
suggesting that strong \CIV\ and \MgII\ absorbers,
on megaparsec scales,
are tracing the distribution of clusters of galaxies,
or of the most luminous super-$L_{*}$ galaxies in those clusters.
The scale of clustering suggests that these absorbers
are biased tracers of the higher--density regions of space,
and that agglomerations of strong absorbers
along a line of sight are indicators of clusters and superclusters.
This is supported by recent observations of Lyman break galaxies.

5.  The growth seen in the clustering of absorbers
is consistent with gravitationally induced growth of perturbations.

\acknowledgments
We wish to acknowledge the long--term direction of Don York,
in compiling the extensive catalog of heavy--element
absorbers used in this study
and in providing intellectual leadership for the project.
We acknowledge helpful discussions 
and useful statistical comments from Carlo Graziani.
JMQ was supported in part by NASA grants GDP 93-08, NAG 5-4406, and NAG 5-2868.
DEVB was supported in part by the Adler Fellowship at the University of
Chicago, by NASA Space Telescope grant GO-06007.01-94A,
and by the Harlan J. Smith Postdoctoral Fellowship.

\appendix
\section{Calculation of the Likelihood Function $\cal L$.}

Recall that the 
correlation function $\xi_{\rm aa}(r,z)$ relates the {\em observed} 
density in the number of absorber pairs, $n(r,z)$, 
separated by comoving distance $r$ at redshift $z$, 
with the {\em expected} density in the number of pairs, $n_0(r,z)$, 
from an unclustered absorber population at the same separation and redshift:
\begin{equation}
{d^2n\over dr dz} = {d^2n_0\over dr dz}\left[ 1 + \xi_{\rm aa}(r,z) \right]\; .
\end{equation}

The {\em likelihood} $\cal L$ is the probability of the data $D$,
given a model (described by a set of parameters $M$) for
the correlation function $\xi_{\rm aa}(r,z)$: ${\cal L} = P(D|M)$.
The data $D$ consist of an observed set of absorber pairs $i$,
each with comoving separation $r_i$ at redshift $z_i$, 
contained in a small cell of size $\delta r_i\,\delta z_i \rightarrow 0$.
We include in the likelihood not only the probability $P_i(1)$
that one pair was observed in the $i$th cell $\delta r_i\,\delta z_i$,
but also the probability $P_k(0)$ that no pair was seen in the remaining
empty cells $\delta r_k\,\delta z_k$:
\begin{eqnarray}
{\cal L} & = & \prod_i P_i(1) \times \prod_{k\neq i} P_k(0)\nonumber \\
& = & \prod_i {d^2n\over dr dz}\Bigg|_{(r_i,z_i)} \delta r_i\,\delta z_i\,
\exp \left(-{d^2n\over dr dz}\Bigg|_{(r_i,z_i)} 
\delta r_i\,\delta z_i\right) \times
\prod_{k\neq i} \exp \left(-{d^2n\over dr dz}\Bigg|_{(r_k,z_k)} 
\delta r_k\,\delta z_k\right)
\nonumber \\
& = & \prod_i {d^2n\over dr dz}\Bigg|_{(r_i,z_i)} \delta r_i\,\delta z_i \times
\prod_{k} \exp \left(-{d^2n\over dr dz}\Bigg|_{(r_k,z_k)} 
\delta r_k\,\delta z_k\right) \; .
\end{eqnarray}
                                                                                                                                                                                                                                                                                                                                                                             
In taking the limit as the cell size goes to zero,
and substituting equation~(A1) into equation~(A2), we find:
\begin{eqnarray}
\log{\cal L} & = &\lim_{\delta r_i\,\delta z_i\to 0}
\sum_i \log\left({d^2n\over dr dz}\Bigg|_{(r_i,z_i)} 
\delta r_i\,\delta z_i\right) 
- \int\int {d^2n\over dr dz}\, dr dz\nonumber \\
 & = & \sum_i \log\left[ 1 + \xi_{\rm aa}(r_i,z_i) \right] - 
\int\int {d^2n_0\over dr dz}\, \xi_{\rm aa}(r,z)\,  dr dz  + {\rm C} \; ,
\end{eqnarray}
where the constant C is just a function of the data 
and has no dependence on the model parameters.
We compute $n_0(r,z)$ and the integral  by generating a large number of
simulated catalogs of unclustered absorbers,
using a bootstrap Monte Carlo method described in \S\ 4 and in QVY.
The integral term is then the mean $\langle \xi_{aa} \rangle_{\rm MC}$
of the model correlation function,
obtained by averaging $\xi_{\rm aa}(r,z)$
over all random pairs in the Monte Carlo simulation,
with $r$ and $z$ falling in the domain of interest:
\begin{equation}
\log{\cal L} = \sum_i \log\left[ 1 + \xi_{\rm aa}(r_i,z_i) \right] - 
\langle \xi_{aa} \rangle_{\rm MC} + {\rm C} \; .
\end{equation}

\newpage

\figcaption[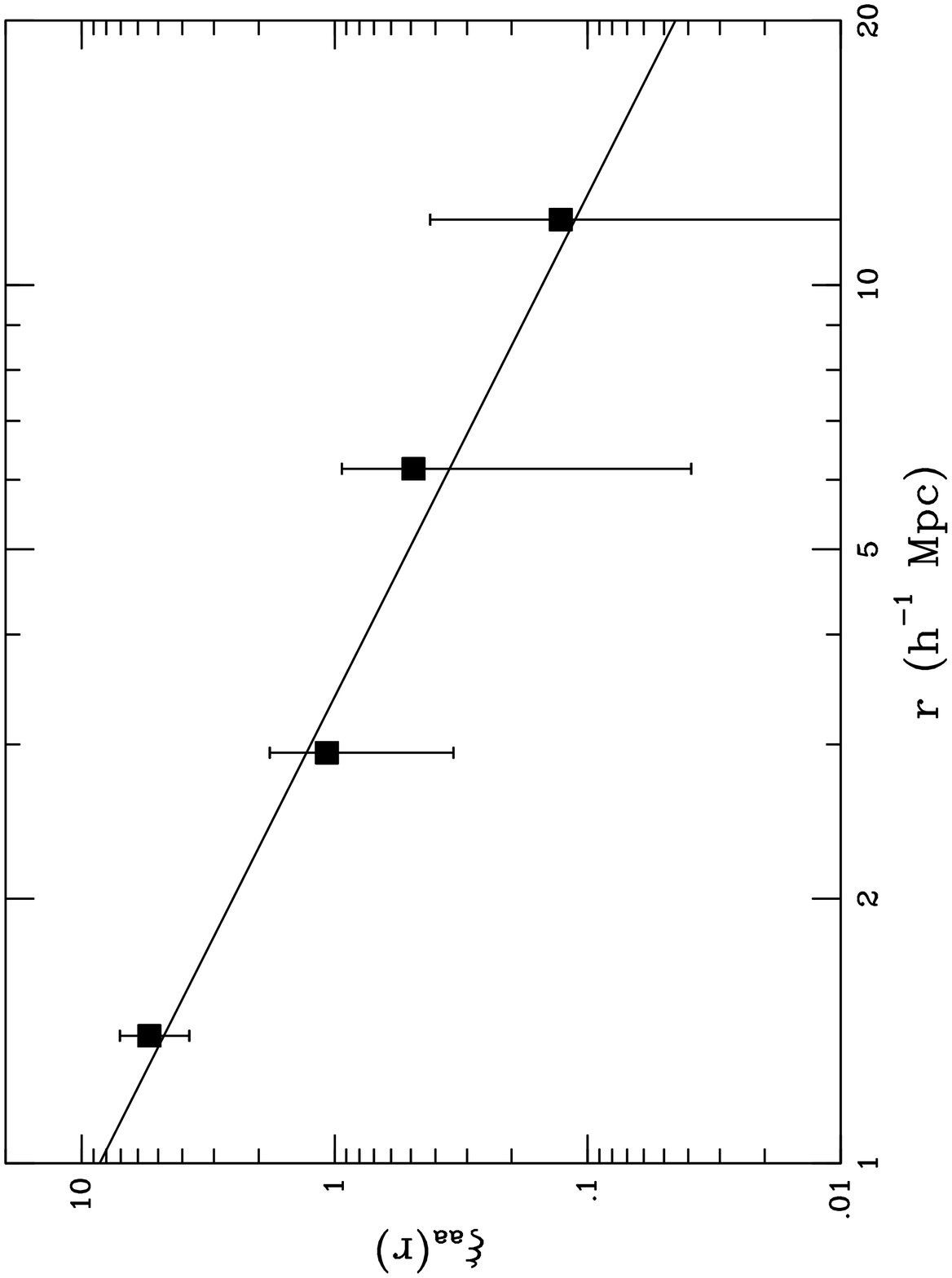]{Line--of--sight correlation function, 
$\xi_{\rm aa}(r)$, for the entire sample of \CIV\ absorbers, 
as a function of absorber comoving separation, $r$, 
in 4 logarithmic bins from 1 to 16 \hMpc .
The vertical error bars through the data points 
are 1 $\sigma$ errors in the  estimator for $\xi_{\rm aa}$.
Also shown is a power--law fit
of the form $\xi_{\rm aa}(r)={\left(r_0/r\right)}^\gamma$,
with maximum--likelihood values $\gamma = 1.75$
and comoving $r_0 = 3.4$ \hMpc\ ($q_0=0.5$). \label{fig1}}

\figcaption[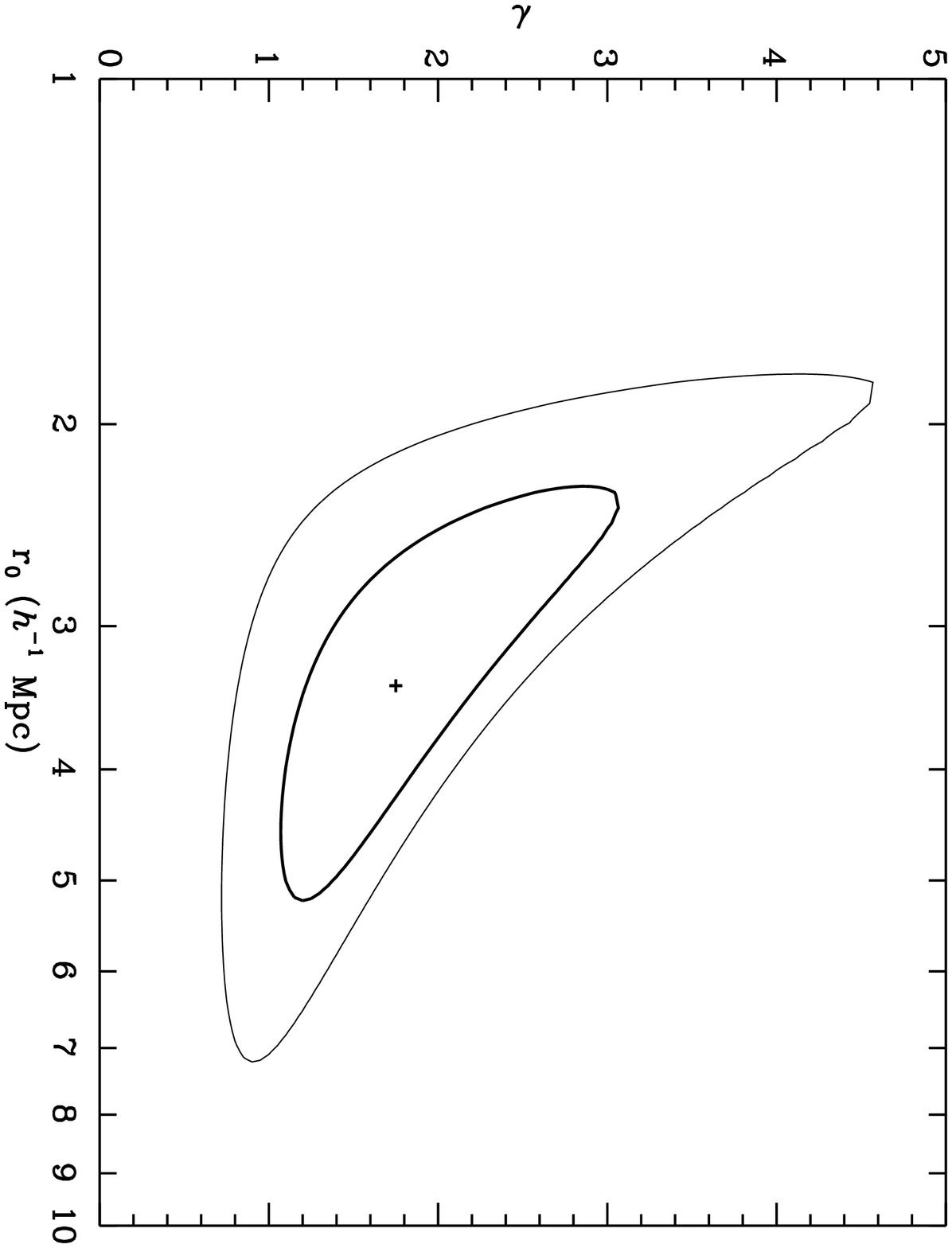]{Credible region for the parameters
in the power--law fit of the form 
$\xi_{\rm aa}(r)={\left(r_0/r\right)}^\gamma$.
The cross marks the maximum--likelihood location, at
$r_0=3.4$ \hMpc\ ($q_0=0.5$) and $\gamma=1.75$.
The heavy and light contours mark the 1 $\sigma$ and 2 $\sigma$
credible regions, respectively. \label{fig2}}

\figcaption[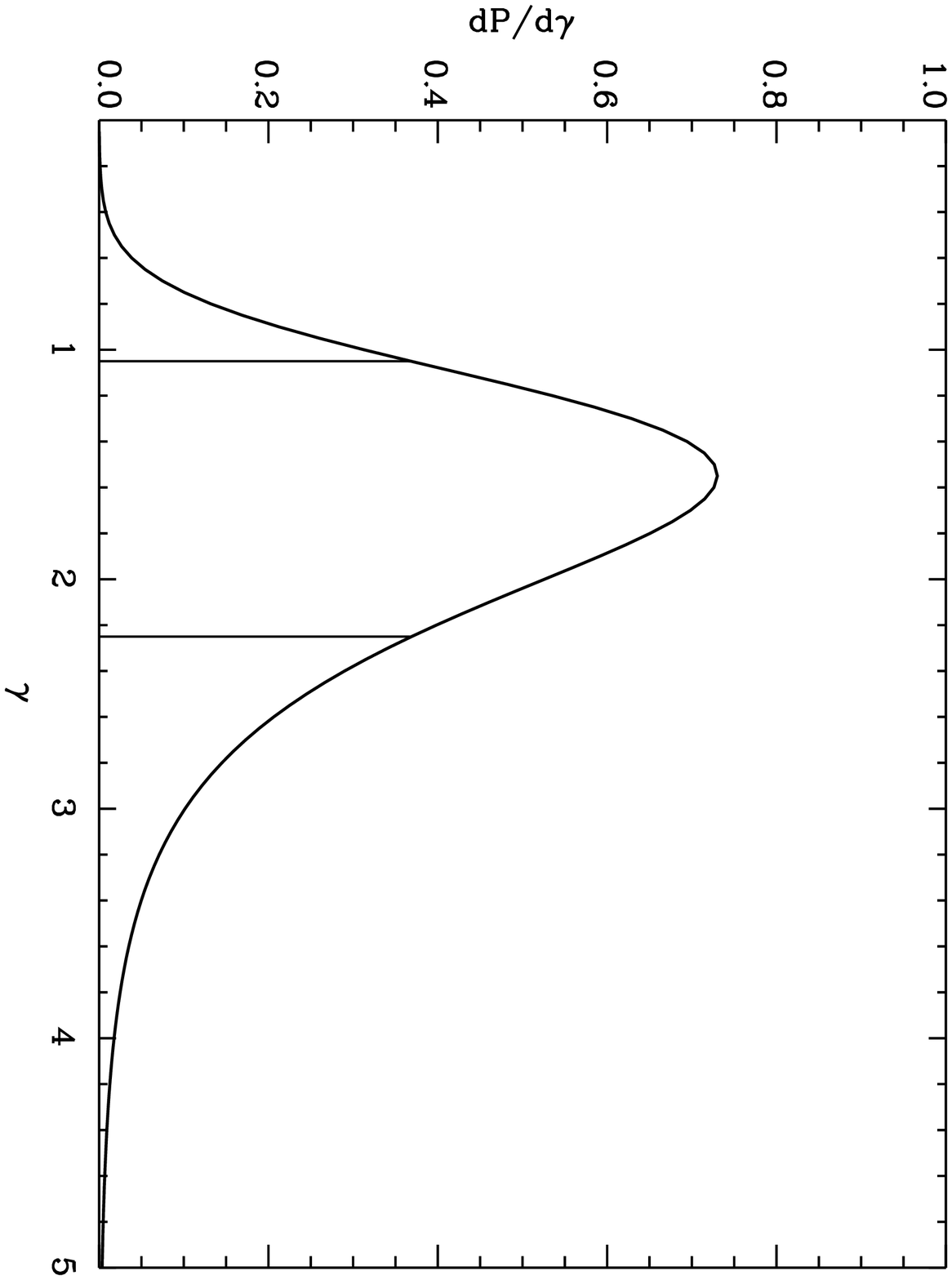]{Posterior distribution for the power--law index
$\gamma$. Vertical lines mark the 1 $\sigma$ credible region,
$1.05 < \gamma < 2.25$. \label{fig3}}

\figcaption[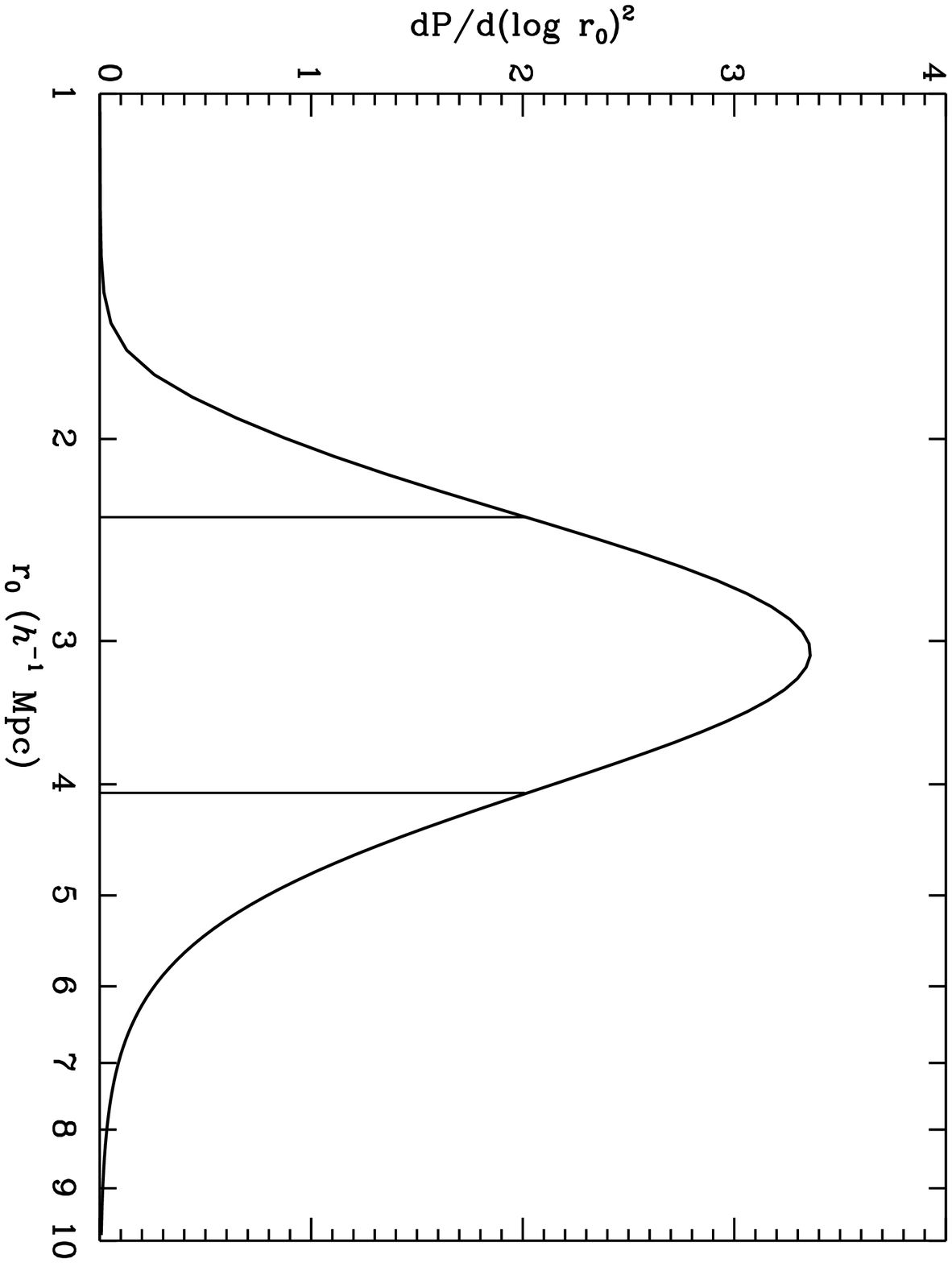]{Posterior distribution for the comoving
correlation length $r_0$. Vertical lines mark the 1 $\sigma$ credible region,
2.4 \hMpc\ $ < r_0  < $ 4.1 \hMpc\ ($q_0=0.5$). \label{fig4}}

\figcaption[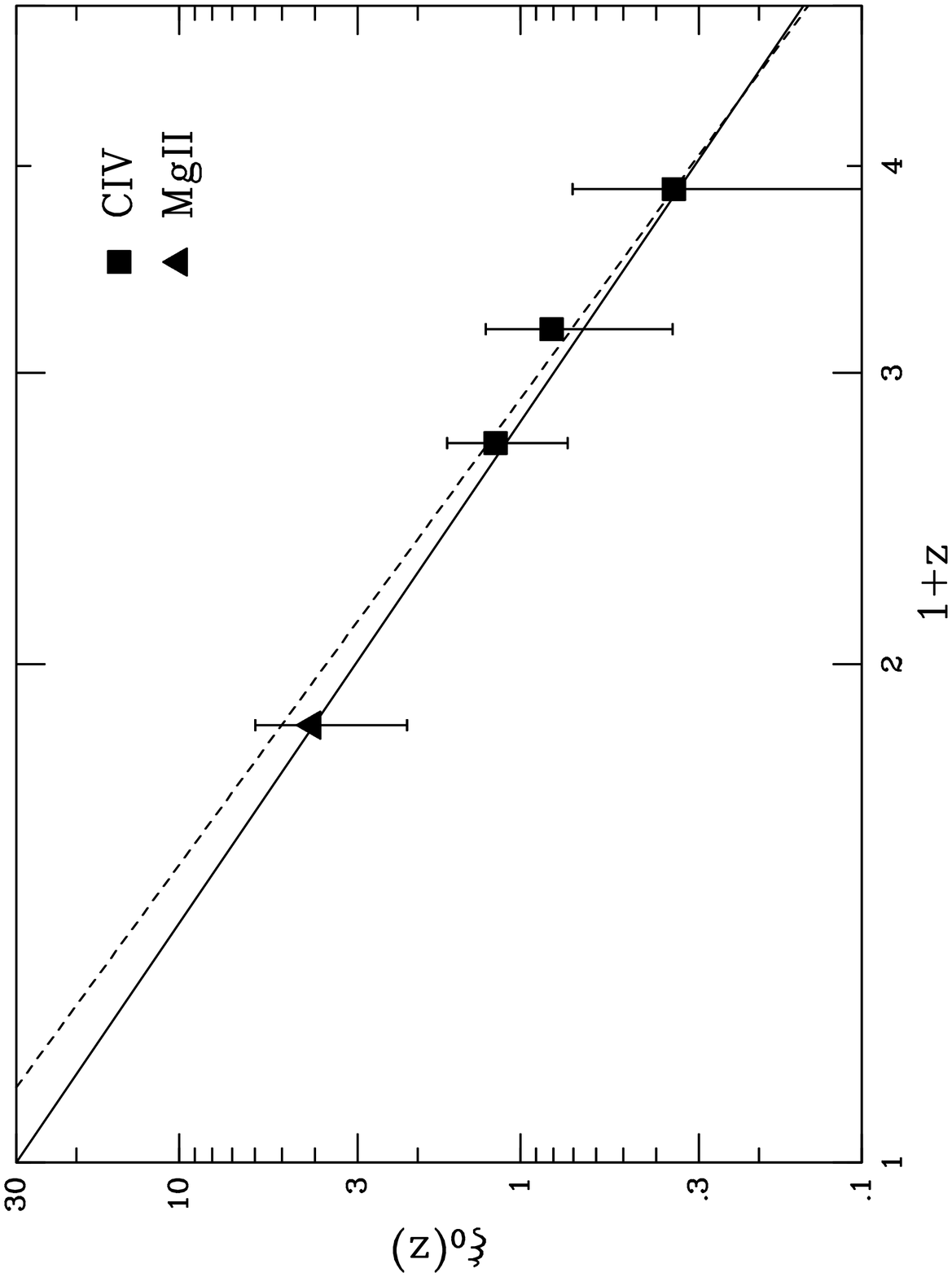]{Mean correlation function,
$\xi_0(z)$, averaged over comoving scales 
$r$ from 1 to 16 \hMpc, as a function of redshift.
Shown are values for the low ($1.2<z<2.0$), medium ($2.0<z<2.8$), 
and high ($2.8<z<3.6$) redshift \CIV\ sub--samples,
as well as for the \MgII\ sample ($0.3<z<1.6$).
The solid line is a maximum--likelihood fit of the form
$\xi_0(z)\propto (1+z)^{-(3+\epsilon)+\gamma}$,
with $\epsilon = 2.05$ and $\gamma=1.75$ ($q_0=0.5$).
The dashed line is the best fit when only the \CIV\ data are used
(see text).}

\figcaption[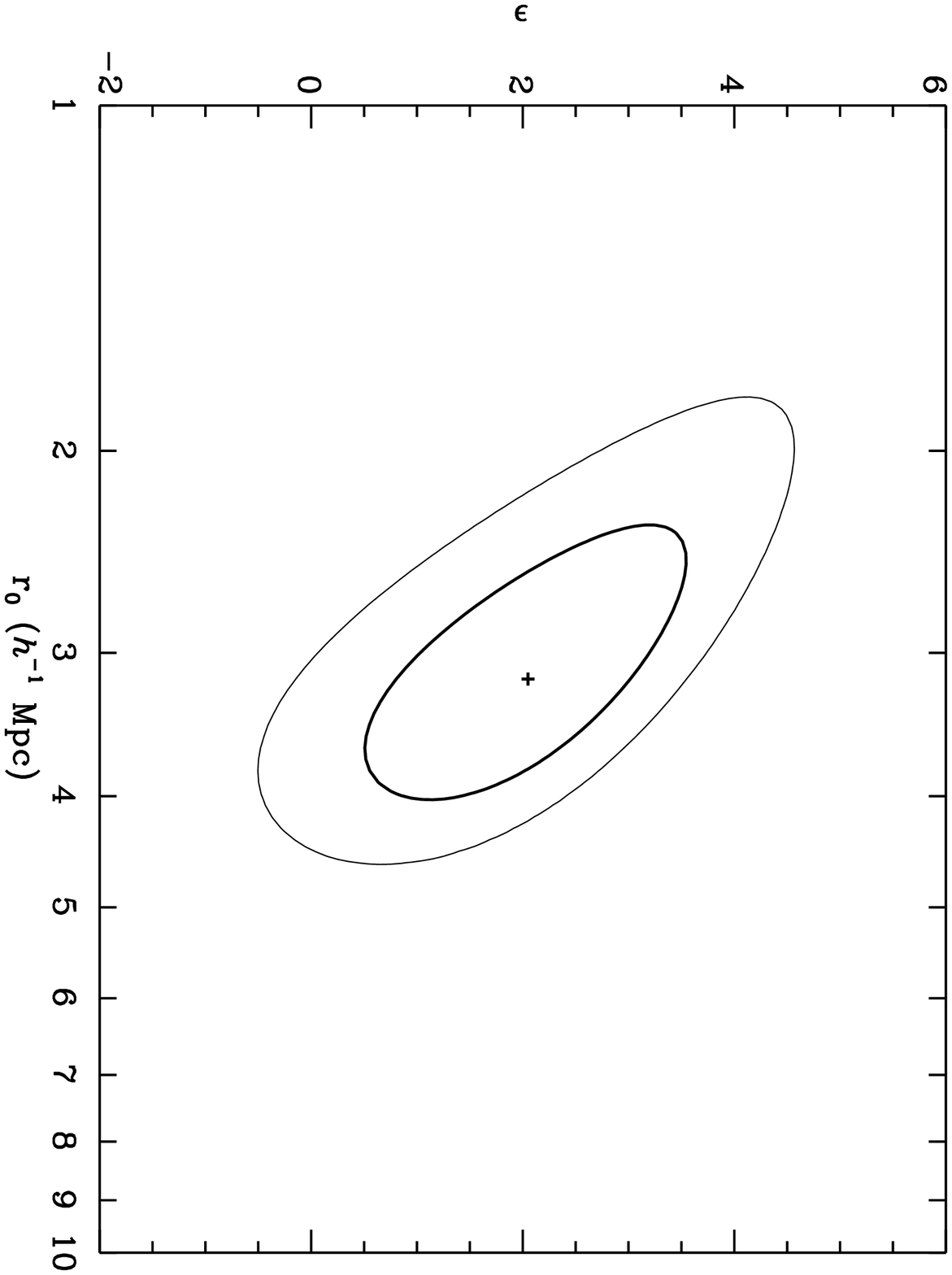]{Credible region for the parameters in the
evolving power--law fit of the form
$\xi_{\rm aa}(r,z) = {\left(r_0/r\right)}^\gamma 
{\left({1+z\over 1+z_0}\right)}^{-(3+\epsilon)+\gamma}$.
We have fixed $\gamma = 1.75$ and $z_0=2.2$, and $q_0=0.5$. 
The cross marks the maximum--likelihood location, 
at $\epsilon=2.05$ and $r_0 = 3.2$ \hMpc.
The heavy and light contours mark the 1 $\sigma$ and 2 $\sigma$
credible regions, respectively. \label{fig6}}

\figcaption[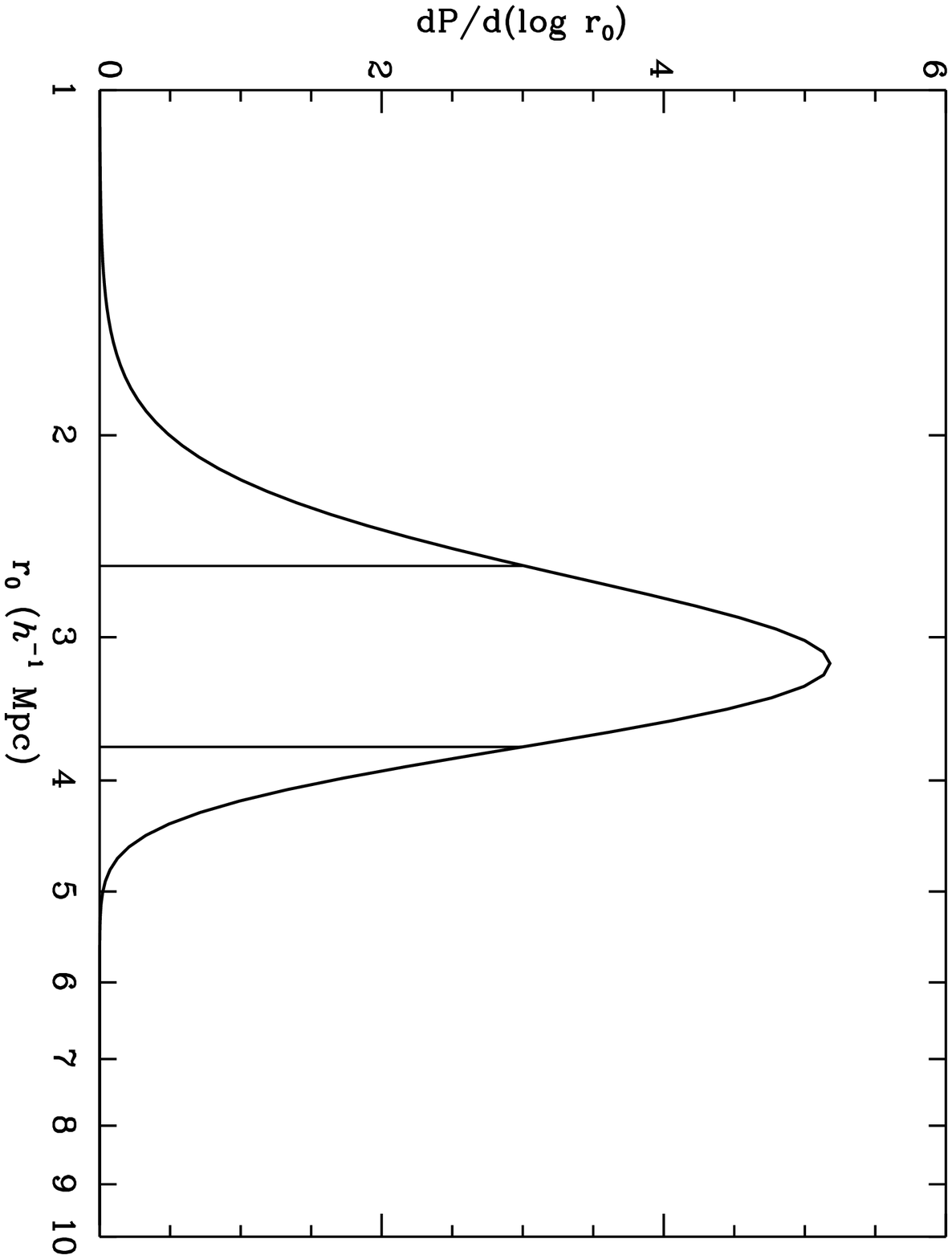]{Posterior distribution for the comoving
correlation length $r_0$ at redshift $z_0=2.2$.
Vertical lines mark the 1 $\sigma$ credible region,
2.6 \hMpc\ $ < r_0 < $  3.8 \hMpc\ ($q_0=0.5$). \label{fig7}}

\figcaption[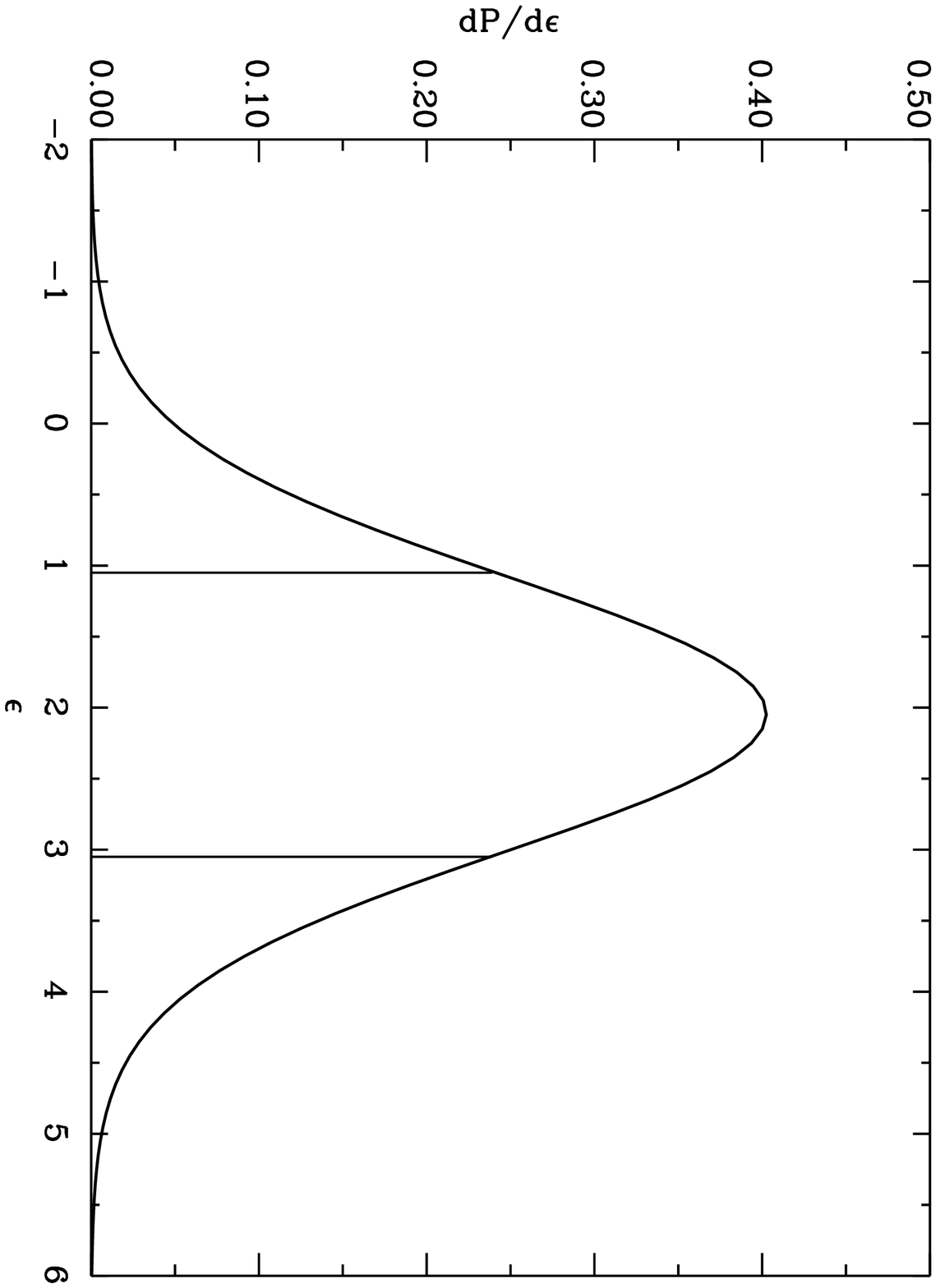]{Posterior distribution for the evolutionary parameter
$\epsilon$. Vertical lines mark the 1 $\sigma$ credible region,
$1.05 < \epsilon < 3.05$ ($q_0=0.5$). \label{fig8}}

\newpage
\begin{figure}
\figurenum{1}
\plotone{fig1.eps}
\caption{}
\end{figure}

\newpage
\begin{figure}
\figurenum{2}
\plotone{fig2.eps}
\caption{}
\end{figure}

\newpage
\begin{figure}
\figurenum{3}
\plotone{fig3.eps}
\caption{}
\end{figure}

\newpage
\begin{figure}
\figurenum{4}
\plotone{fig4.eps}
\caption{}
\end{figure}

\newpage
\begin{figure}
\figurenum{5}
\plotone{fig5.eps}
\caption{}
\end{figure}

\newpage
\begin{figure}
\figurenum{6}
\plotone{fig6.eps}
\caption{}
\end{figure}

\newpage
\begin{figure}
\figurenum{7}
\plotone{fig7.eps}
\caption{}
\end{figure}

\newpage
\begin{figure}
\figurenum{8}
\plotone{fig8.eps}
\caption{}
\end{figure}

\end{document}